\documentclass[aps,superscriptaddress,twocolumn]{revtex4-2}

\usepackage{amsmath}    % need for subequations
\usepackage{graphicx}   % need for figures
\usepackage{verbatim}   % useful for program listings
\usepackage{xcolor}      % use if color is used in text
\usepackage{subfigure}  % use for side-by-side figures
\usepackage{hyperref}   % use for hypertext links, including those to external documents 
\usepackage{natbib}
\usepackage{float}
\usepackage{chngcntr} %for appendix counter
\newcommand{\bo}[1]{\textbf{#1}}

\begin{document}

%\title{Effect of Ge-substitution on Magnetic Properties in the itinerant chiral magnet MnSi
\title{Direct observations of spin fluctuations in spin-hedgehog-anti-hedgehog lattice states in MnSi$_{1-x}$Ge$_x$ ($x=0.6$ and $0.8$) at zero magnetic field}
\author{Seno Aji}
\email{senji77@issp.u-tokyo.ac.jp}
\affiliation{Institute for Solid State Physics (ISSP), The University of Tokyo, Kashiwa 277-0882, Japan.}
\author{Tatsuro Oda}
\affiliation{Institute for Solid State Physics (ISSP), The University of Tokyo, Kashiwa 277-0882, Japan.}
\author{Yukako Fujishiro}
\affiliation{RIKEN Center for Emergent Matter Science (CEMS), Saitama 351-0198, Japan.}
\author{Naoya Kanazawa}
%\affiliation{Department of Applied Physics and Quantum Phase Electronics Center (QPEC), The University of Tokyo, Tokyo 113-8656, Japan.}
\affiliation{Institute of Industrial Science, The University of Tokyo, 4-6-1 Komaba Meguro-ku, Tokyo 153-8505, Japan.}
\author{Hiraku Saito}
\affiliation{Institute for Solid State Physics (ISSP), The University of Tokyo, Kashiwa 277-0882, Japan.}
\author{Hitoshi Endo}
\affiliation{Institute of Materials Structure Science, High Energy Accelerator Research Organization, Tsukuba, Ibaraki 305-0801, Japan.}
\affiliation{Materials and Life Science Experimental Facility, J-PARC Center, Tokai, 319-1195, Japan.}
\author{Masahiro Hino}
\affiliation{Institute for Integrated Radiation and Nuclear Science, Kyoto University, Kumatori, Osaka 590-0494, Japan.}
\author{Shinichi Itoh}
\affiliation{Institute of Materials Structure Science, High Energy Accelerator Research Organization, Tsukuba, Ibaraki 305-0801, Japan.}
\affiliation{Materials and Life Science Experimental Facility, J-PARC Center, Tokai, 319-1195, Japan.}
\author{Taka-hisa Arima}
\affiliation{RIKEN Center for Emergent Matter Science (CEMS), Saitama 351-0198, Japan.}
\affiliation{Department of Advanced Materials Science, The University of Tokyo, Kashiwa 277-8561, Japan.}
\author{Yoshinori Tokura}
\affiliation{RIKEN Center for Emergent Matter Science (CEMS), Saitama 351-0198, Japan.}
\affiliation{Department of Applied Physics and Quantum Phase Electronics Center (QPEC), The University of Tokyo, Tokyo 113-8656, Japan.}
\affiliation{Tokyo College, The University of Tokyo, Tokyo 113-8656, Japan.}
\author{Taro Nakajima}
\email{taro.nakajima@issp.u-tokyo.ac.jp}
\affiliation{Institute for Solid State Physics (ISSP), The University of Tokyo, Kashiwa 277-0882, Japan.}
\affiliation{RIKEN Center for Emergent Matter Science (CEMS), Saitama 351-0198, Japan.}

%\date{12 January 2019}

\begin{abstract}

The helimagnetic compounds MnSi$_{1-x}$Ge$_{x}$ show the three-dimensional multiple-$q$ order as referred to as spin-hedgehog-anti-hedgehog (SHAH) lattice. Two representative forms of SHAH are cubic-3$q$ lattice with $q \| \langle100\rangle$ and tetrahedral-4$q$ lattice with  $q \| \langle111\rangle$, which show up typically for $x=1.0-~0.8$ and for $x=0.6$, respectively.  Here, we have investigated the spin fluctuations in the MnSi$_{1-x}$Ge$_{x}$ polycrystalline samples with $x=0.6$ and $0.8$ by using the time-of-flight (TOF) neutron inelastic scattering and MIEZE-type neutron spin echo techniques to elucidate the microscopic origin of the unconventional Hall effect in the SHAH lattice states. This research is motivated by the observation of a sign change in the unconventional Hall resistivity as a function of temperature [Y. Fujishiro et al., Nat. Comm. \bo{10}, 1059 (2019)]. The present results reveal the correspondences between the temperature ranges where the positive Hall resistivity and spin fluctuations are observed. These results agree well with the theoretical model of the conduction electrons scattered by the fluctuating spin clusters with a non-zero average of sign-biased scalar spin chirality as a mechanism of the positive Hall resistivity [H. Ishizuka and N. Nagaosa, Sci. Adv. \bo{4}, eaap9962 (2018)].

\end{abstract}

\maketitle

\section{Introduction}

\begin{figure*}[t]
\centering
\includegraphics[width=14 cm]{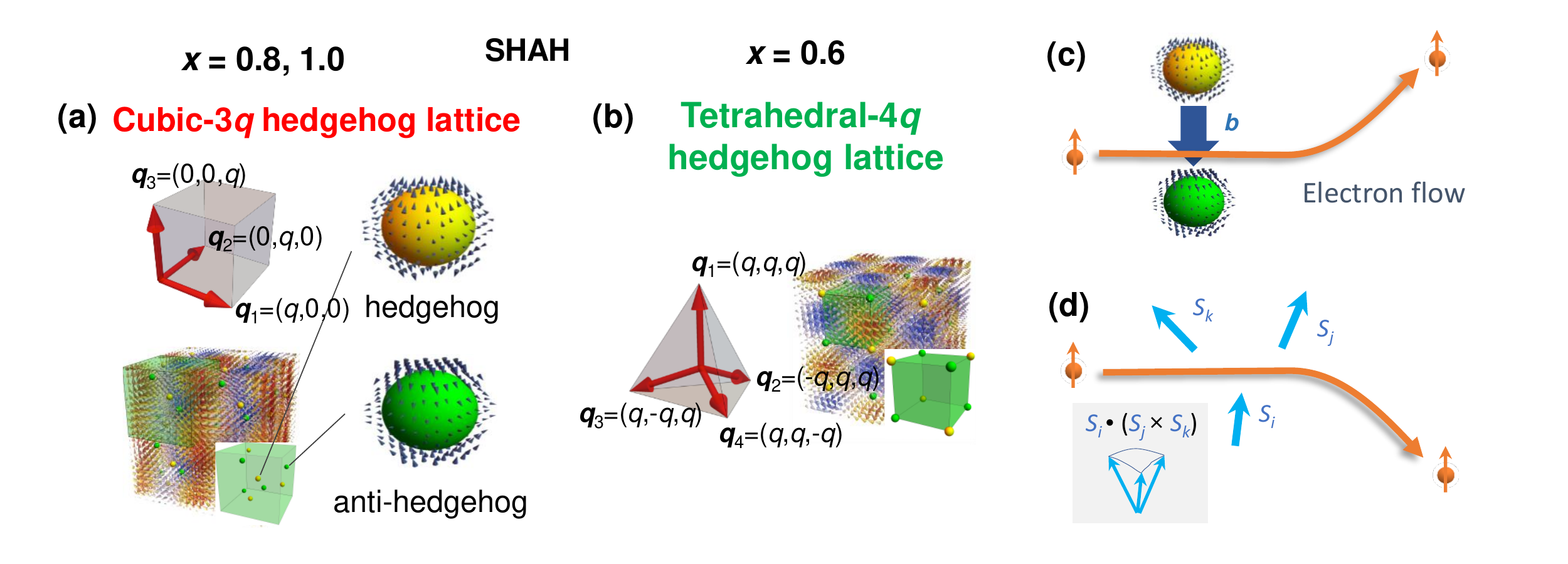}
\caption{Spin configurations in the (a) cubic-3$q$ and (b) tetrahedral-4$q$ lattice states. Schematics showing (c) a conduction electron deflected by an effective magnetic field $b$ which is generated by the spin hedghog and anti-hedghog, and (d) the skew scattering process due to the spin-cluster with scalar spin chirality.}
\label{fig:Fig01_final}
\end{figure*}

The interplay between conduction electrons and non-coplanar magnetism is a topic of growing interest in condensed matter physics \cite{PhysRevLett.83.3737,PhysRevLett.87.116801,PhysRevLett.98.057203,doi:10.1126/science.1058161,RevModPhys.82.1539}. Non-coplanar magnetic orders give rise to an anomalous Hall effect of unconventional origin. One important example showing a large Hall effect is the pyrochlore ferromagnet Nd$_2$Mo$_2$O$_7$, in which the Nd and Mo spins exhibit non-coplanar orders~\cite{doi:10.1126/science.1058161}. A conduction electron hopping over three sites with noncoplanarly arranged spins of $S_i$, $S_j$ and $S_k$ acquires a Berry phase \cite{berry}, which is proportional to the scalar spin chirality $S_i \cdot (S_j \times S_k)$ \cite{doi:10.1143/JPSJ.73.2624,PhysRevLett.93.096806}. %
The Berry phase can be regarded as an effective magnetic field acting on the conduction electrons, and induces an unconventional Hall effect, which is proportional to neither the external magnetic field nor magnetization. % 
Another example is the Hall effect owing to the magnetic skyrmion lattice phase. %
A magnetic skyrmion~\cite{Sov.Phys.JETP.68.101,BOGDANOV1994255,Rossler2006}, which was first discovered in a noncentrosymmetric cubic helimagnet MnSi, is a topologically non-trivial vortex-like spin object where the sign of the scalar spin chirality of three neighboring magnetic moments is fixed \cite{doi:10.1126/science.1166767,Nagaosa2013}.  % 
A previous Hall resistivity measurement on the archetypal skyrmion host compound MnSi demonstrated that an additional Hall resistivity appears only in the skyrmion lattice phase \cite{PhysRevLett.102.186602}. 

The topological Hall effect was also found in a short-period helimagnet MnGe \cite{PhysRevLett.106.156603}. %
This material is isostructural to MnSi which belongs to the cubic $B$20 compounds of space group $P2_{1}3$, but remarkably exhibits a different type of topological spin order. %
The magnetic structure is determined to be a triple-$q$ state with $q$-vectors of ($q$,0,0), (0,$q$,0) and (0,0,$q$), which results in a spin-hedgehog-anti-hedgehog (SHAH) lattice state; hereafter we refer to as $3q$-SHAH state, as shown in Fig.~\ref{fig:Fig01_final}(a) \cite{PhysRevB.86.134425,Tanigaki2015}. %
By calculating the effective magnetic field $b$~arising from the scalar spin chirality [Fig.~\ref{fig:Fig01_final}(c)], it turns out that the SHAH lattice state has an equal number of effective magnetic monopoles and antimonopoles, which are cancelled out at zero external magnetic field. %
However, the application of an external magnetic field leads to the displacement of the monopoles and antimonopoles, which induces a net effective magnetic field evidenced by a large topological Hall effect \cite{Kanazawa2016}. %
In fact, the field evolution of the Hall resistivity of MnGe at low temperatures is successfully reproduced by calculating the field dependence of the scalar spin chirality. %

There remain several unsolved problems in the topological Hall effect in MnGe. %
One is the sign reversal of the Hall resistivity near the critical temperature. %
The sign of the topological Hall effect is determined by the direction of the effective magnetic field and coupling between the conduction electrons and localized magnetic moments, which are supposed to be independent of temperature~\cite{PhysRevLett.93.096806}. %
%Nevertheless, the Hall resistivity in MnGe changes from negative to positive near the critical temperature~\cite{PhysRevLett.106.156603}. %
Nevertheless, the Hall resistivity in MnGe changes from positive to negative as the temperature is lowered below the critical temperature~\cite{PhysRevLett.106.156603}.
A recent theoretical study suggested that the positive Hall resistivity can be explained by skew scattering due to fluctuating but locally correlated spins with a non-zero average of sign-biased scalar spin chirality [Fig.~\ref{fig:Fig01_final}(d)] \cite{doi:10.1126/sciadv.aap9962}. %
To corroborate this scenario, we need to quantitatively investigate how the spin fluctuations develop with varying temperature. %
There was a neutron resonance spin-echo study on MnGe reporting spin fluctuations near the $Q$-position where a magnetic Bragg peak develops \cite{PhysRevB.99.100402}. %
However, the analysis of the $Q$-dependence of the spin fluctuations and the comparison with the transport data are still lacking. %

Another problem is the doping dependence revealed by recent neutron scattering and Hall resistivity measurements \cite{Fujishiro2019}. %
By substituting Ge for Si, particularly in the composition range of $0.3 < x < 0.7$, the magnetic structure of MnSi$_{1-x}$Ge$_{x}$ changes to quadrupole-$q$ magnetic order described by a superposition of four equivalent $q$-vectors of $(q,q,q), (-q,q,q), (q,-q,q)$ and $(q,q,-q)$, referred to as  $4q$-SHAH state, as shown in Fig.~\ref{fig:Fig01_final}(b). For~$x<0.3$, it turns into the single-$q$ helical state, similar to MnSi.
The $4q$-SHAH state also shows large unconventional Hall effect and the sign changes not only near the critical temperature but also at low temperatures. %
The correlation between the Hall resistivity and spin fluctuations in the $4q$-SHAH state has not been investigated thus far. %

In the present study, we investigate the spin fluctuations of polycrystalline MnSi$_{1-x}$Ge$_{x}$ samples with $x=~0.6$ and $0.8$, which exhibit the $4q$- and $3q$-SHAH lattice states, respectively, by means of the time-of-flight neutron inelastic scattering and the neutron resonance spin-echo spectroscopy. %
In both samples, we found that the spin fluctuations develop in a wide range of temperatures centered around the critical temperatures $T_\mathrm{c}$, in which the positive Hall resistivity appears. %
%However, we also found that the positive Hall resistivity observed at the low temperature regions in the $x=0.6$ sample cannot be explained by the spin fluctuations scenario, implying that there is another microscopic mechanism for the sign reversal in the $4q$-SHAH state. %

\section{Experimental Details}
Polycrystalline samples of MnSi$_{1-x}$Ge$_{x}$ were prepared by the high-pressure synthesis technique. Mn, Si, and Ge chunks were first mixed with the stoichiometric
ratio and then melted in an arc furnace under an argon atmosphere. Afterwards, it was heated at 1073 K for 1 h under 5.5–6.0 GPa with a cubic-anvil type
high-pressure apparatus. Powder x-ray diffraction analyses confirmed B20-type crystal structure ($P2_{1}3$). The samples were sealed in an Al-cell with $^4$He gas for the neutron experiments.

%A time-of-flight (TOF) neutron inelastic scattering experiment using a high-resolution chopper spectrometer with a pulsed white neutron beam was performed at the HRC beamline (BL12)~ \cite{ITOH201190} 
%Time-of-flight (TOF) neutron inelastic scattering experiments were performed at the high-resolution chopper spectrometer HRC (BL12)~\cite{ITOH201190} in the materials and life science experimental facility (MLF) of J-PARC, %

Time-of-flight (TOF) neutron inelastic scattering experiments were performed at the High-Resolution Chopper spectrometer (HRC) at BL12~\cite{ITOH201190} in the Materials and Life science experimental Facility (MLF) of J-PARC, %
in order to obtain the magnetic scattering intensity distribution in the wide $Q-E$ space with varying temperature. The Fermi chopper A \cite{ITOH201976} %
with a frequency of 100 Hz and the target incident energy of $E_\mathrm{i}$= 3.5 meV were selected to optimize the energy resolution. %Fe collimator 30-30 was used with the incident collimator 0.3 deg-long. 
The beam size at the sample position was approximately $30\times30$ mm$^2$. The horizontal beam collimation was 0.3 degrees. 
The energy resolution at the elastic condition was obtained as $0.089$~meV.

\begin{figure*}[t]
\centering
\includegraphics[width=16.5 cm]{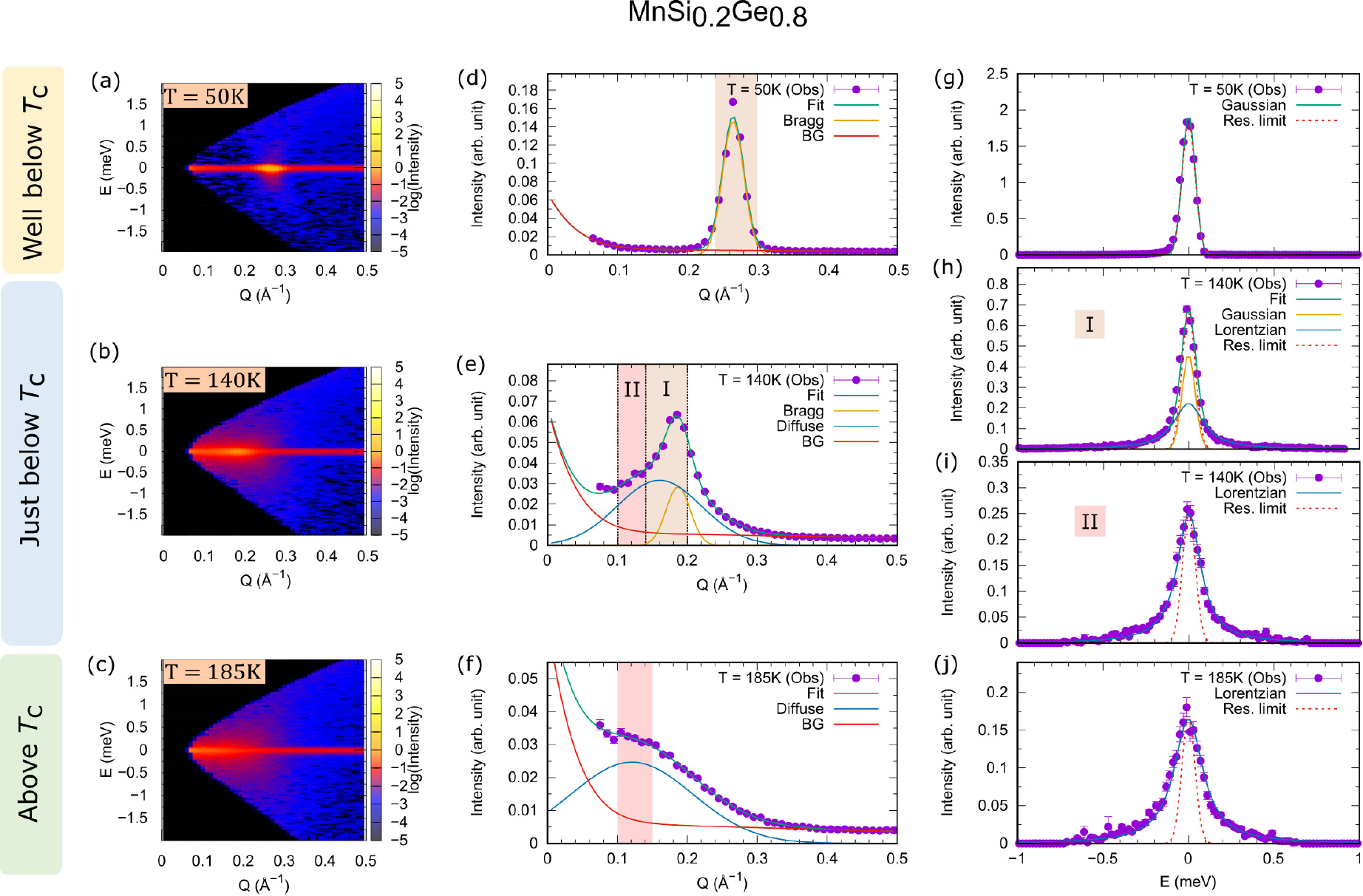}
\caption{Neutron inelastic scattering results for Ge-concentration of $x = 0.8$ at three representative temperatures. (a)-(c) Intensity maps in the logarithmic scales as a function of momentum ($Q$) and energy ($E$), (d)-(f) constant-$E$ profiles with the integration of energy range of $-1<E<1$ meV derived from (a)-(c), and (g)-(j) constant-$Q$ profiles with the integration of $Q$-ranges are displayed in the color area of (d)-(f). The critical temperature ($T_\mathrm{c}$) in this sample is obtained as $150$~K.
% intensity as a function of $Q$ at $E=[-1.1]$, and (h)-(k) inte
%, (d)-(g) and (h)-(k) are the intensity profiles for const-$E$ and const-$Q$ cuts, respectively, derived from (a)-(c).
}
\label{fig:HRC_result_x0p8}
\end{figure*}

%The investigation of the property of the spin fluctuations was performed by MIEZE-type Neutron Spin Echo (NSE) measurements at the VIN ROSE beamline (BL06)
We also investigated the spin fluctuations by MIEZE-type Neutron Spin Echo (NSE) measurements at the VIN ROSE beamline (BL06)~\cite{MasahiroHino,PhysRevApplied.14.054032} in MLF of J-PARC, which provides pulsed polychromatic beam. A spin-polarized incident neutron beam was obtained by supermirror polarizers. A weak magnetic field of approximately $0.5$ mT was applied in the beam path from the polarizers to a spin analyzer to maintain the spin polarization. A pair of resonance spin flippers of RSF1 ($f_1$) and RSF2 ($f_2$) was used to control probabilities of flipping the neutron spins in a wide range of the neutron wavelengths. Both RSF1 and RSF2 acted as $\pi/2$ flippers to give the flipping probabilities of $50\%$ for all the wavelengths. 
After passing through RSF1 and RSF2, the wave function of the incident neutron beam was composed of four states with different energies. Finally, two of them with spin-down states were removed by a supermirror spin analyzer. As a result, the wave function of incident beam at the sample position was the superposition of the two spin-up states with the energy difference corresponding to the modulation frequency, $f_M = f_1 - f_2$ \cite{GAHLER1992899,doi:10.1063/1.4965835}. The modulation frequency was set to $f_M=10$~kHz in the present experiment. The incident neutron beam at the sample position has a size of approximately $5\times 10$~mm$^2$. The scattered neutrons were detected by a two dimensional-sensitive detector.

In MIEZE-type NSE experiments, the Fourier time $t$ of the intermediate scattering fucntion $I(Q,t)$ is given by

\begin{equation}
t=(\frac{m}{h})^2f_ML_{sd}\lambda^3
\end{equation}

\noindent where $m$ and $h$ are the neutron mass and the Planck constant, respectively. $L_{sd}$ is a distance between the sample and the detector, which was $0.325$ m in the present experiment. $\lambda$ is the neutron wavelength, which is determined from the time of flight (TOF) of each neutron. Substituting the wavelengths range of the incident neutron beam at VIN ROSE, which spans from 3.2 to 11.5 \r{A}, into Eq. (1), the Fourier time range is deduced to be from 0.6 to 32~ps. 

The intermediate scattering function of $I(Q,t)$ profiles were obtained from the ratio of the contrasts of the intensity modulations between the scattered and incident beam for the specified TOF following the method given in Ref.~\cite{PhysRevResearch.2.043393}. The elastic scattering yields the same contrast, while the inelastic one leads to a reduction of the contrast due to the adiabatic change in the velocity of the scattered neutron. This TOF-MIEZE method enables us to measure the $I (Q, t )$ values with different sets of $Q$ and $t$ simultaneously \cite{doi:10.1063/1.4965835}.

\begin{figure*}[t]
\centering
\includegraphics[width=13.5 cm]{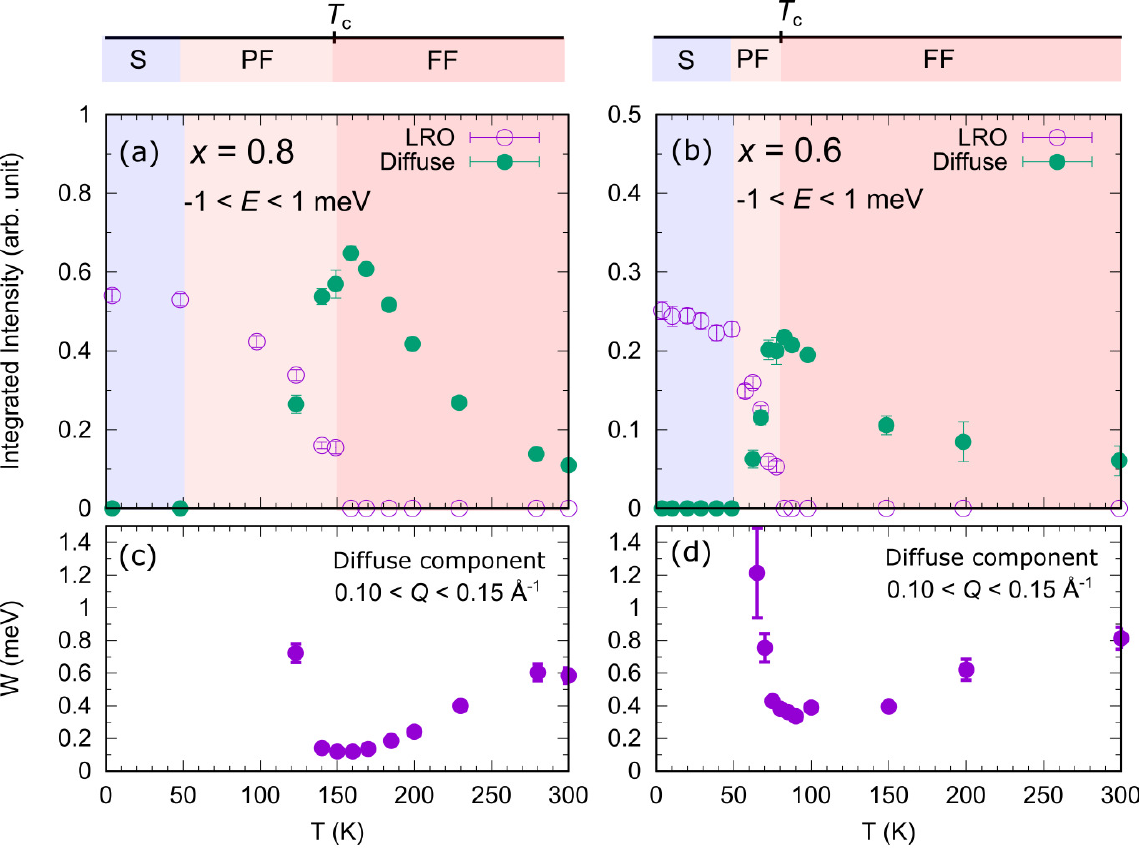}
\caption{Integrated intensities (a)-(b) and intrinsic widths $W$ (c)-(d) as a function of temperature for $x = 0.8$ (left) and $x = 0.6$ (right), respectively. Integrated intensities are derived from the intensity profiles as a function of $Q$ or constant-$E$ profiles near the elastic conditions, the LRO and diffuse components can be deconvoluted by two Gaussian functions. The static (S), partially fluctuating (PF), and fully fluctuating (FF) regions can be distinguished. Meanwhile, the intrinsic widths are deduced from the constant-$Q$ profiles of the diffuse components at lower $Q$ regions with the $Q$-ranges of $0.10 < Q < 0.15$ \r{A}$^{-1}$.  }
\label{fig:HRC_result_Int_x0p6_x0p8_width_eps}
\end{figure*}

\section{Results and Discussions}
%\subsection{Neutron inelastic scattering measurements using a high-resolution chopper spectrometer}
%\subsection{High-Resolution Chopper Spectrometer}
%\subsection{Time-of-flight neutron inelastic scattering using a high-resolution chopper spectrometer}
\subsection{TOF Neutron Inelastic Scattering}

To grasp an overview of the temperature dependence of the magnetic scattering intensity distributed in the $Q-E$ space, 
%we first performed the time-of-flight neutron inelastic scattering experiment using a high-resolution chopper spectrometer at HRC. %
we first performed the TOF neutron inelastic scattering experiment at HRC.
In Figs. \ref{fig:HRC_result_x0p8} (a)-\ref{fig:HRC_result_x0p8}(c), we show the intensity maps of the $x =0.8$ sample at three representative temperatures of $T=50, 140,$ and $185$~K. %
At $50$~K, the magnetic Bragg peak is clearly observed at around $q_\mathrm{m} \sim 0.26$ \r{A}$^{-1}$. %
As we increase the temperature to 140 K, which is still lower than the critical temperature of this sample ($T_\mathrm{c}=150$ K), the magnetic Bragg peak moves to the lower $Q$-region ($q_\mathrm{m} \sim 0.19$ \r{A}$^{-1}$). In addition, the diffuse scattering significantly develops in the low-Q region, and the diffuse intensity is spreading along both $Q$- and $E$-directions. %
At $T=185$~K, the Bragg peak completely disappears, but the strong diffuse scattering still remains. 

To quantitatively investigate the widths and peak positions of the Bragg and diffuse scattering, we cut the intensity maps into profiles with respect to $Q$ and $E$; the former and the latter are referred to as constant-$E$ and constant-$Q$ profiles, respectively. %
Figure \ref{fig:HRC_result_x0p8}(d) shows the constant-$E$ profile near the elastic condition in which the intensities in the energy range of $-1 < E < 1$ meV are integrated, measured at 50 K. %
We observe a sharp peak corresponding to the magnetic long-range order (LRO). % 
The constant-$Q$ profile of the Bragg peak also shows the resolution-limited sharp profile, as shown in Fig. \ref{fig:HRC_result_x0p8}(g). %
These results indicate that there is no spin fluctuations except for possible collective magnon excitations, which were not clearly observed in the present experiment due to the lack of statistics. %

The diffuse scattering at 140 K is clearly seen in the constant-$E$ profile shown in Fig.~\ref{fig:HRC_result_x0p8}(e). % 
Additional intensity emerges in the low-$Q$ region besides the Bragg peak. %
By fitting two Gaussian functions to the data, the additional component is described by a broad peak located at around $Q=0.16$ \AA$^{-1}$, which is lower than the position of the Bragg peak. %
Figures \ref{fig:HRC_result_x0p8}(h) and \ref{fig:HRC_result_x0p8}(i) show the constant-$Q$ profiles in the $Q$ regions labelled "I" and "II" in Fig.~\ref{fig:HRC_result_x0p8}(e), respectively; %
%the former contains the intensities of the Bragg and broad peaks while the latter contains only those from the broad peak. %
both the Bragg and broad peaks contribute to the former, while the latter comprises only of the broad peak. %
The profile in the region I is well reproduced by a summation of a resolution-limited Gaussian and a broad Lorentzian functions. %
By contrast, the profile in the region II is fitted just by a Lorentzian function. %
The broad peak in the constant-$E$ profile is also observed above $T_\mathrm{c}$, and it also has a Lorentzian shape as a function of $E$, as shown in Figs.  \ref{fig:HRC_result_x0p8}(f) and \ref{fig:HRC_result_x0p8}(j), respectively. %
These data demonstrated that the broad peak emerging in the low-$Q$ region correspond to the diffusive spin fluctuations. %
Note that the diffuse scattering near $T_\mathrm{c}$ was also reported in previous small-angle neutron scattering (SANS) studies of MnGe\cite{PhysRevB.86.134425,PhysRevLett.125.137202}. %
The present data show that the $x=0.8$ sample exhibits a similar spin dynamics to that of pure MnGe. %

\begin{figure*}[t]
\centering
\includegraphics[width=14 cm]{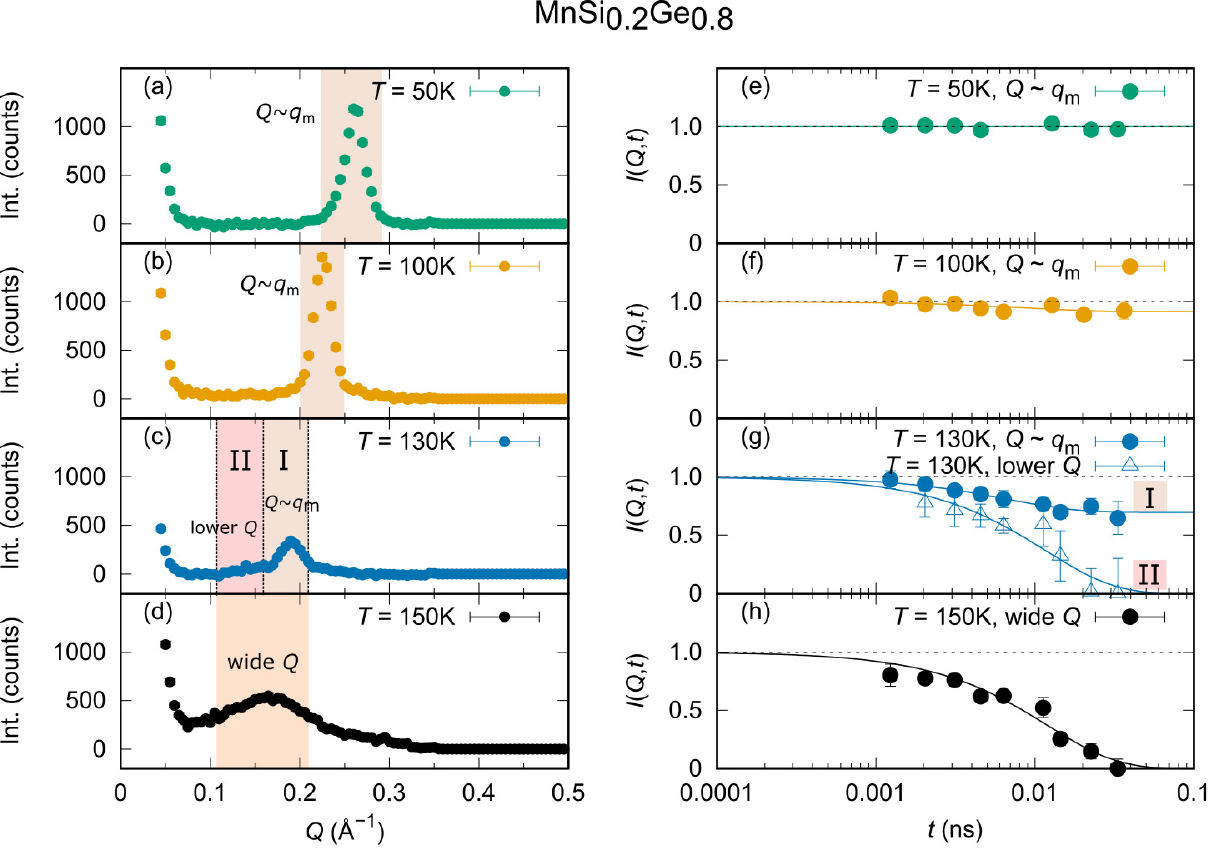}
\caption{MIEZE-NSE experimental results for $x = 0.8$. (a)-(d) $Q$-dependence of intensity at four representative temperatures, and (e)-(h) the intermediate scattering function of $I(Q,t)$ deduced from the selected $Q$-range given in (a)-(d).}
\label{fig:Iqt-comparison_x0p8}
\end{figure*}

We also performed a neutron inelastic scattering experiment on the $x=0.6$ sample in the same manner as that for the $x=0.8$ sample, as shown in Appendix~\ref{AP:one}. %
In Figs. \ref{fig:HRC_result_Int_x0p6_x0p8_width_eps}(a)-\ref{fig:HRC_result_Int_x0p6_x0p8_width_eps}(b), we summarize the temperature dependence of the integrated intensities of the Bragg and diffuse scatterings derived from constant-$E$ profiles near the elastic conditions as well as the instrinsic widths of the diffuse scattering at lower $Q$ regions ($0.10 < Q < 0.15$ \r{A}$^{-1}$), the intrinsic width is approximated as ${W} = \sqrt{W_\mathrm{obs}^2 -W_\mathrm{res}^2}$, where $W_\mathrm{obs}$ and $W_\mathrm{res}$ are the observed and resolution-limited widths, respectively. %
As shown in Figs. \ref{fig:HRC_result_Int_x0p6_x0p8_width_eps}(a)-(b), we can see that both samples show the similar behavior i.e. the LRO components are realized at low temperatures, the LRO and diffuse components coexist at the middle range of temperatures, and only diffuse components exist at high temperatures. The static (S), partially fluctuating (PF), and fully fluctuating (FF) regions for both samples can be separated. %
The temperature evolution of the intrinsic widths [Figs. \ref{fig:HRC_result_Int_x0p6_x0p8_width_eps}(c)-(d)] shows that the diffuse components undergo the critical slowing down at the critical temperature $T_\mathrm{c}$. %where the phase transition occurs. %
%
%Note that, in the $x=0.8$ ($x=0.6$) sample, the constant-$E$ profile at $T=100$~K ($T=60$~K) was slightly broader than the profiles at low temperatures. However, it was difficult to resolve the spin fluctuations from the constant-$Q$ profile because the intensity of the diffuse and Bragg components were significantly reduced and enhanced, respectively, as we lowered temperature. These points will be investigated further by the MIEZE-type neutron spin echo in the following section.

As the Bragg peaks grow with decreasing temperature, the observed intensities are dominated by the static components of the magnetic moments in the systems, and thus the fractions and characteristic time of the fluctuating spin components become rather difficult to evaluate at low temperatures. %
Specifically, in the $x = 0.8$ ($x = 0.6$) sample, the constant-$E$ profile at $T = 100$~K ($T = 60$~K) is slightly broader than the profiles at low temperatures. Hence, it is difficult to resolve the spin fluctuations from the constant-$Q$ profile. To unambiguously determine the temperature dependence of the fraction of the fluctuating spin components, we thus performed the MIEZE- type neutron spin echo spectroscopy for both the samples.

 \subsection{MIEZE-type Neutron Spin Echo}

\begin{figure*}[t]
\centering
\includegraphics[width=17 cm]{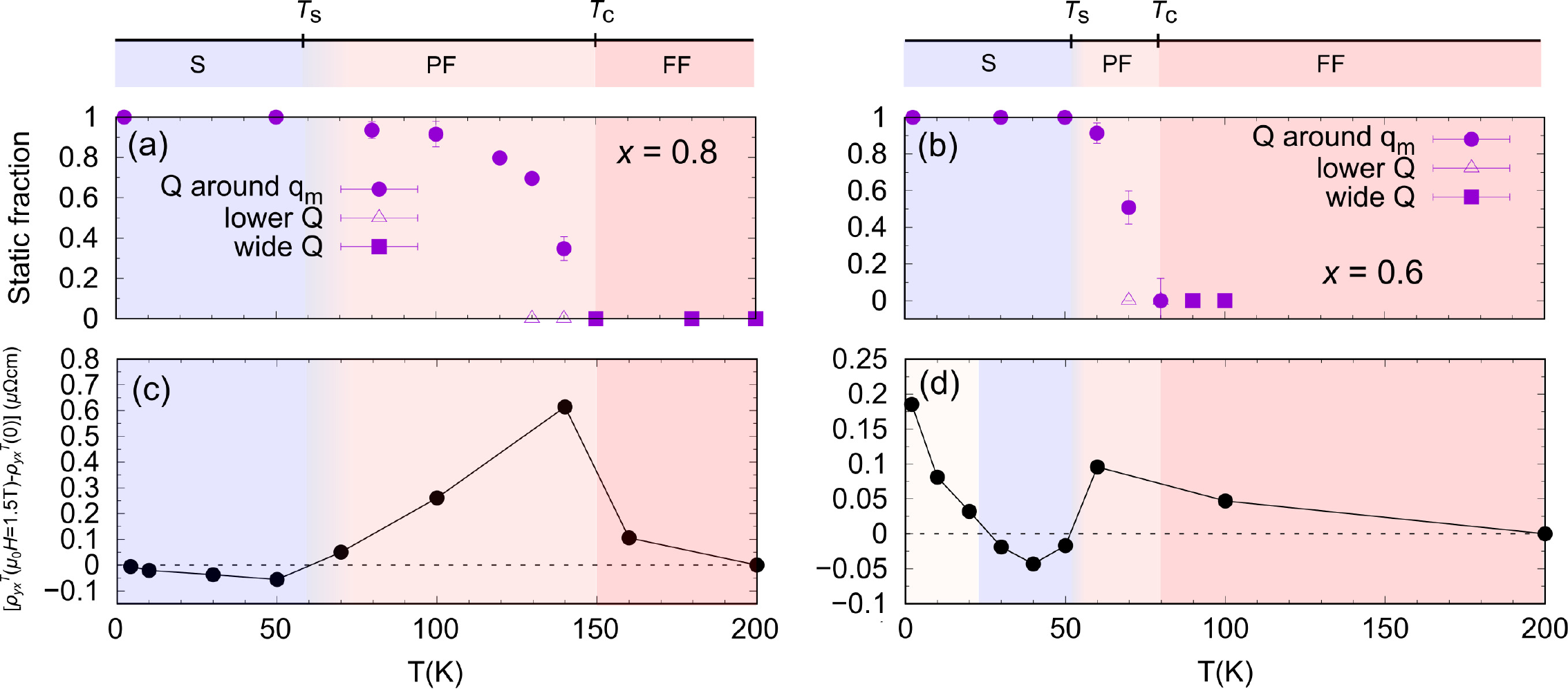}
\caption{The correspondences between the spin fluctuations and positive Hall resistivity for $x=0.8$ (left) and $x=0.6$ (right), respectively. Temperature dependence of the static fraction (a)-(b), and the topological Hall resistivity at $\mu_0H=1.5$~T, $[\rho_{yx}^{T}(\mu_0 H=1.5\mathrm{T})-\rho_{yx}^{T}(0)]$ (c)-(d). The static fractions are derived from the present MIEZE-NSE measurements, while the topological Hall resistivity data were extracted from Ref. \cite{Fujishiro2019}.}
\label{fig:S0tauvsTemp_x0p8_x0p6_ver02}
\end{figure*}

Figures \ref{fig:Iqt-comparison_x0p8}(a)-\ref{fig:Iqt-comparison_x0p8}(d) are the intensity profiles as a function of momentum $Q$ in the $x=0.8$ sample at four representative temperatures of $T=50, 100, 130,$ and $150$~K measured at VIN ROSE. % and selected time of flight, $\rm{TOF}=28$~ms. %
These profiles were obtained by extracting the intensities measured by $\lambda = 5.2 \pm 0.3$~\r{A} from the polychromatic TOF scattering data. %
%The data are basically the same as those measured at HRC; the well-defined peaks attributed to the long-range orders are observed at low temperature and get broadened at high temperature. 
Similar to the results at HRC, we observed the well-defined peaks attributed to the long-range orders at low temperature and get broadened at high temperature.
To distinguish the contributions from the static and fluctuating components, the integration ranges of $Q$ are selected for each representative temperatures. From the selected integration ranges of $Q$, the intensity profiles of $I(Q,t)$ as a function of the Fourier time $t$ are acquired [Figs.~\ref{fig:Iqt-comparison_x0p8}(e)-~\ref{fig:Iqt-comparison_x0p8}(h)]. %
To extract the static/fluctuating fractions and the characteristic time of the fluctuation, we performed fitting analysis using an exponential decay function and a constant term, 

\begin{equation}
I(Q,t)=S_0 + (1-S_0)e^{-t/\tau}, 
\end{equation}
\noindent where $S_0$ and $\tau$ are the static fraction and the characteristic time of the spin fluctuations, respectively.

At low temperature of $T=50$ K [Fig.~\ref{fig:Iqt-comparison_x0p8}(e)], the observed peak is clearly static (S) since the $I(Q,t)$ profile is flat in time domain where the static fraction $S_0=~1$ (no fluctuating spins at all).
In the middle range of temperatures, we observe the coexistence of the static and fluctuating components or partially fluctuating (PF). 
At $T=130$ K, we deduce two $I(Q,t)$ functions corresponding to the $Q$-ranges labeled I and II, revealing that the $I(Q,t)$ function in the region I remains finite in the $t\rightarrow\infty$ limit, while the $I(Q,t)$ function in the region II decays to zero. %
The system becomes fully fluctuating (FF) when the temperature reaches $150$~K [Fig. \ref{fig:Iqt-comparison_x0p8}(h)] indicated by the decaying curve without the static fraction. %
These observations are consistent with the HRC results. %
Importantly, the $I(Q,t)$ function at 100 K slightly deviates from unity. This result unambiguously shows that the spin fluctuations still remain at this temperature. %

\begin{figure*}[t]
\centering
\includegraphics[width=15 cm]{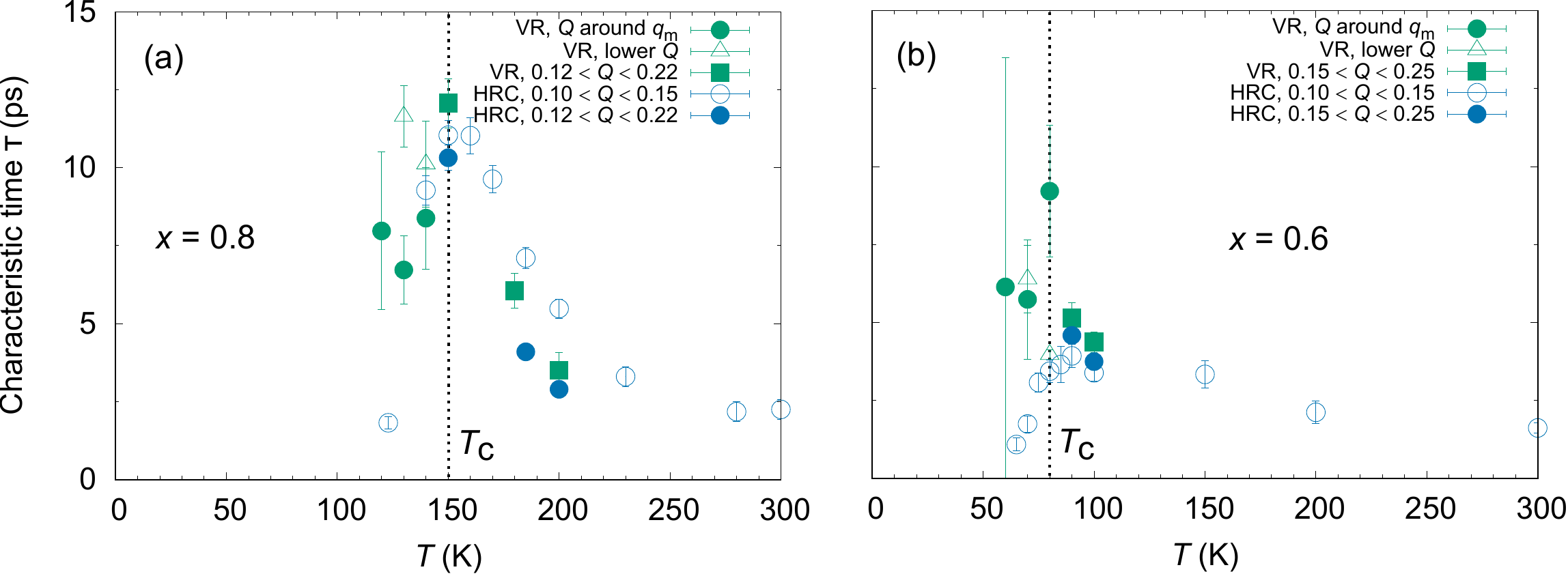}
\caption{Characteristic time of spin fluctuations $\tau$ as a function of temperature for (a) $x=0.8$ and (b) $x=0.6$. The data are deduced from the MIEZE-NSE experiments (VR) at $Q$ around $q_\mathrm{m}$ (Bragg) and lower-$Q$ (Diffuse). The $Q$-ranges are in the unit of \r{A}$^{-1}$. The approximate characteristic times of the diffuse components from the time-of-flight neutron inelastic experiments (HRC) are also present.}
\label{fig:TauvsTemp_x0p6_x0p8}
\end{figure*}

We also performed MIEZE-type neutron spin echo measurements on the $x=0.6$ sample, as shown in Appendix \ref{AP:two}. %
The static fractions as a function of temperature for both $x=0.8$ and $x=0.6$ samples are summarized in Figs. \ref{fig:S0tauvsTemp_x0p8_x0p6_ver02}(a)-(b). It is shown clearly that the fluctuating spins start to develop around $T_{\rm{s}}=60$~K and $T_{\rm{s}}=50$~K for the $x=0.8$ and $x=0.6$ samples, respectively. The magnetic long-range order and the fluctuating spins coexist in the middle ranges of temperatures of $T_{\rm{s}}<T<T_{\rm{c}}$, where the critical temperatures  ($T_{\rm{c}}$) are obtained as $150$~K and $80$~K for the Ge-concentrations of $x=0.8$ and $x=0.6$, respectively. These results support the experimental data measured at HRC shown in Fig.~\ref{fig:HRC_result_Int_x0p6_x0p8_width_eps}. 

We show the temperature dependence of the topological Hall resistivity at $\mu_0H=1.5$~T (much lower than the critical field) in order to see the sign reversal behavior [Figs. \ref{fig:S0tauvsTemp_x0p8_x0p6_ver02}(c)-(d)]. %
%We show the correspondences between the positive sign of topological Hall resistivity [Figs. \ref{fig:S0tauvsTemp_x0p8_x0p6_ver02}(a)-(b)] and spin fluctuations [Figs. \ref{fig:S0tauvsTemp_x0p8_x0p6_ver02}(c)-(d)]. %
By comparing with Figs. \ref{fig:S0tauvsTemp_x0p8_x0p6_ver02}(a)-(b), we observed the spin fluctuations in a wide range of temperature centered at $T_\mathrm{c}$, which coincides with the temperature range where the positive Hall resistivity was observed. %
These results confirm the theoretical model that the conduction electrons are scattered by the fluctuating spin clusters with a non-zero average of sign-biased scalar spin chirality \cite{doi:10.1126/sciadv.aap9962}. %
As mentioned in the introduction, the $x=0.6$ sample exhibits positive Hall resistivity not only near the critical temperature but also at low temperatures. %
The present results clearly show that there is no spin fluctuations at the lowest temperature in the $x=0.6$ sample, indicating that the fluctuating spin cluster model cannot be applied to the low-temperature regime of the $x=0.6$ sample. %
Note that, in MnSi$_{1-x}$Ge$_x$ compound, the isovalent substitution of Si for Ge atoms does not change the number of charge carrier thus the electronic Fermi level remains unchanged. The Si-substitution will result in change of the lattice constant due to the chemical pressure and also the magnetic modulation length \cite{Fujishiro2019,PhysRevMaterials.3.104408}, which may modify the Fermi surface due to the nesting-type instability of the magnetic modulation vector-$q$. A sign change of the topological Hall effect by the Fermi surface modification at low temperatures had been studied rigorously in Mn$_{1-x}$Fe$_x$Si system \cite{PhysRevLett.112.186601}. It is challenging for the theory side to confirm this interesting scenario in MnSi$_{1-x}$Ge$_x$, as well. This scenario is beyond our experimental study.

We also found that in both samples, the diffuse scattering still remains even at 300 K as shown in Fig. 2, while the topological Hall resistivity disappears above 200 K. %[comment] We should state the experimental results first, and then interpretation for them.
In general, in the paramagnetic phase, the spin correlation length and the characteristic time of the spin fluctuation become shorter with increasing temperature. %
Therefore, the disappearance of the topological Hall resistivity could be interpreted that the spin fluctuations above 200 K no longer contain the spin cluster with a non-zero average of sign-biased scalar spin chirality. %
We note here that the temperature dependence of the vector spin chirality, $S_i\times S_j$, in the paramagnetic phase of MnSi was studied by polarized and unpolarized neutron scattering \cite{PhysRevLett.102.197202}. %
Just above the critical temperature, MnSi exhibits diffuse scattering arising from the short-range spin correlation, in which the sign of $S_i\times S_j$ is completely fixed. %
As the temperature is increased, the chiral spin correlation rapidly decays, and thus the system exhibits non-chiral spin fluctuations at high temperatures. %
The temperature evolution agrees with the energy scale of Dzyaloshinskii-Moriya (DM) interaction in MnSi derived from the neutron scattering experiment of about $1$~meV \cite{grigoriev2009, PhysRevB.96.214425}. We suggest that the scalar spin chirality in the paramagnetic phase of MnSi$_{1-x}$Ge$_x$ also has a similar tendency with respect to temperature. %
Furthermore, the magnitude of DM interaction in MnGe varies depending on the evaluation, the energy scale is less than~$\sim 10$~meV \cite{Koretsune.JPSJ.87.041011,Gayles.PhysRevLett.115.036602}. Thus, it is rather reasonable to suppose that the chiral fluctuations in MnSi$_{1-x}$Ge$_x$ disappear above $\sim 200$~K.
  
%Thus, the disappearance of the chiral fluctuations in MnSi$_{1-x}$Ge$_x$ above $\sim 200$ K is quite reasonable.

%We suggest that the scalar spin chirality in the paramagnetic phase of MnSi$_{1-x}$Ge$_x$ also has a similar tendency with respect to the temperature. %
%Furthermore, in terms of the energy scale of Dzyaloshinskii-Moriya (DM) interaction which contributes to the fluctuating spin clusters, the energy scale of DM interaction in MnSi derived from the neutron scattering experiment is about $1$~meV \cite{grigoriev2009, PhysRevB.96.214425}, while the magnitude of DM interaction in MnGe varies depending on the evaluation approaches and the energy scale is less than~$\sim 10$~meV \cite{Koretsune.JPSJ.87.041011,Gayles.PhysRevLett.115.036602}. Thus, the disappearance of the chiral fluctuations in MnSi$_{1-x}$Ge$_x$ above $\sim 200$ K is quite reasonable.

Next, we also estimated the temperature dependence of the characteristic times of spin fluctuations $\tau$ which can be extracted directly from the MIEZE-NSE measurements [Figs. \ref{fig:TauvsTemp_x0p6_x0p8}(a)-\ref{fig:TauvsTemp_x0p6_x0p8}(b)], it was found in the order of picosecond (ps). We also present the approximate characteristic times $\tau$ derived from the obtained intrinsic energy widths in the HRC experiment, $\tau =\hbar/\gamma$, where $\gamma$ is the half of the intrinsic width $W$ (see Figs. \ref{fig:HRC_result_Int_x0p6_x0p8_width_eps}(c)-(d)). Although the intrinsic width $W$ is only quite roughly estimated from HRC experiment, it is in good agreement with the characteristic time $\tau$ deduced from the MIEZE-NSE measurement (VR).  

The obtained the characteristic times of spin fluctuations~$\tau$ near $T_\mathrm{c}$ in the $x=0.8$ and $x=0.6$ samples are 12 and 9 ps, respectively. %
These characteristic times are sufficiently longer than the time scale of the electrons hopping among the atomic sites, which is typically estimated to be in the order of femtosecond from the Fermi velocity and inter-atomic distances \cite{PhysRevB.90.144404}. Therefore, the scalar spin chirality in the locally correlated spin-cluster can persists during the scattering process of the electrons. 

The present doped samples have shorter characteristic time compared to the pristine MnSi and MnGe. The characteristic times in MnSi and MnGe near $T_{\mathrm{c}}$ are 1~ns \cite{PhysRevLett.102.197202,PhysRevLett.119.047203,PhysRevB.81.144413,PhysRevX.9.041059,PhysRevResearch.2.043393} and 20~ps \cite{PhysRevB.99.100402}, respectively. These results suggest that the characteristic time of the spin fluctuations as a function of $x$ shows a non-monotonous behavior. This characteristic time may be related to the magnetic modulation length since such a non-linear behavior was also observed in the magnetic modulation length as given in Ref. \cite{Fujishiro2019}. Please note that, at low concentration $x$, the magnetic modulation length is well described by the ratio of the conventional symmetric exchange to DM-type antisymmetric exchange interactions~\cite{PhysRevMaterials.3.104408}.~However, those interactions failed to explain the magnetic structure at higher concentrations ($x>0.25$) where the SHAH lattice states stabilized~\cite{Fujishiro2019}. The Ruderman-Kittel-Kasuya-Yosida (RKKY) interaction is considered to be the origin of the SHAH states~\cite{PhysRevB.94.174403,PhysRevB.95.224424}. 

\section{Summary}
In summary, we have studied the spin fluctuations in the representative MnSi$_{1-x}$Ge$_{x}$ polycrystalline samples, i.e.,  the $x=0.8$ compound with cubic-3$q$ SHAH lattice and the $x=0.6$ compound with tetrahedral-4$q$ SHAH lattice, by means of the time-of-flight neutron inelastic scattering and the neutron resonance spin-echo spectroscopy. The present results show that these samples exhibit relatively large spin fluctuations centered at the transition temperatures $T_\mathrm{c}$, and also reveal the correspondences between the temperature ranges where the positive Hall resistivity and spin fluctuations are observed. They agree very well with each other for the whole temperature range in the $x=0.8$ sample, but show a discrepancy at low temperatures region in the $x=0.6$ sample. This discrepancy cannot be explained by the fluctuating spin-cluster mechanism. Another scenario such as the Fermi-surface nesting-type instability need to be addressed for the future work to confirm the low-temperature sign reversal of the Hall resistivity. We suggest that the chiral fluctuations may be vanished above $200$ K in the paramagnetic phase where the diffuse scattering still remains. In addition, we found the coexistence of the magnetic long-range order and spin fluctuations in the middle temperature ranges, $T_\mathrm{s}<T<T_\mathrm{c}$. Just below $T_\mathrm{c}$, we observed the indication of the $Q$-dependence of the characteristic time~$\tau$. Furthermore, the obtained characteristic times in the present doped samples are faster than that in the pristine MnSi and MnGe. This characteristic times may be related to the magnetic modulation length. 

\begin{center}
\bo{ACKNOWLEDGMENTS}
\end{center}

We thank Dr. H. Ishizuka for fruitful discussions. The neutron experiment at the Materials and Life Science Experimental Facility of J-PARC was performed under the user program %(Proposal Nos. 2023S01, 2019S07, 2020B0424, ). 
(Proposal Nos.~2021S01, 2022S01, 2019S07, and 2020B0424).
This work was supported by JSPS KAKENHI (Grants Nos.~19H01856, 22K14011, 22K18965 and 23H04017) and JST FOREST (Grant no.~JPMJFR2038).

\appendix
\counterwithin{figure}{section}

\section{HRC experimental results at x = 0.6}
\label{AP:one}

\begin{figure*}[t]
\centering
\includegraphics[width=17 cm]{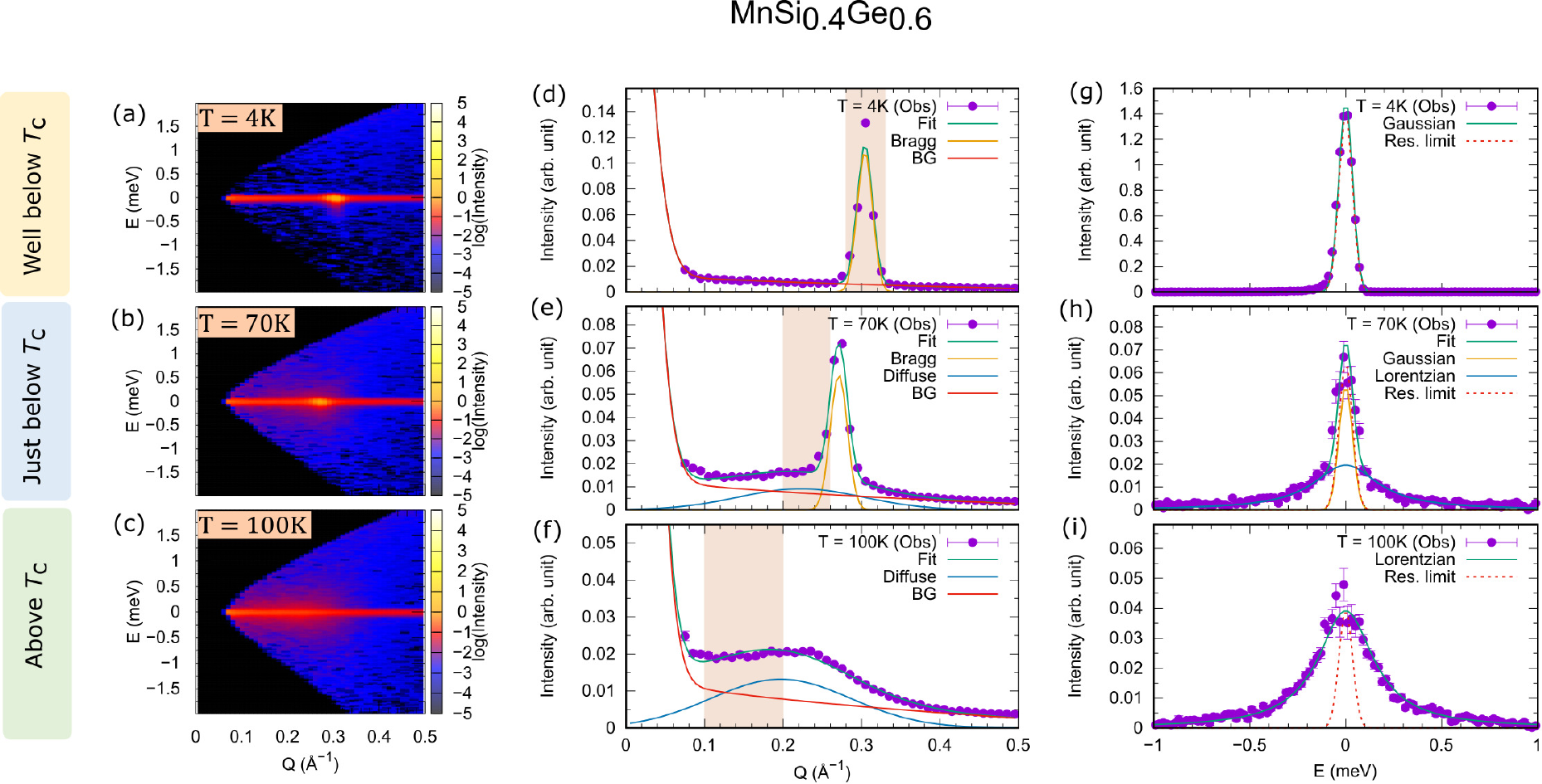}
\caption{Neutron inelastic scattering results for Ge concentration $x = 0.6$ at three representative temperatures. (a)-(c) Logarithmic scales of the intensity maps as a function of momentum ($Q$) and energy ($E$), (d)-(f) constant-$E$ profiles near the elastic condition derived from (a)-(c), and (g)-(i) constant-$Q$ profiles with the $Q$-ranges are displayed in the color area of (d)-(f). The critical temperature ($T_\mathrm{c}$) in this sample is about $ 80$ K.}
\label{fig:HRC_result_x0p6}
\end{figure*}

We show the time-of-flight neutron inelastic experiment in the $x=0.6$ sample. At low temperature of $T=4$~K, the magnetic Bragg peak is clearly observed in the intensity map [Fig. \ref{fig:HRC_result_x0p6} (a)]. The constant-$E$ profile with the energy range of $-1<E<1$ meV shows a sharp peak corresponding to the magnetic long-range order (LRO) at around $q_\mathrm{m} \sim 0.30$ \r{A}$^{-1}$ [Fig. \ref{fig:HRC_result_x0p6}(d)]. By integrating over the $Q$-range around the Bragg peaks, the constant-$Q$ profile can be obtained, as shown in Fig. \ref{fig:HRC_result_x0p6}(g). We found that the obtained constant-$Q$ profile coincides well with the resolution-limited Gaussian profile indicating that the spin fluctuations are not present. As we increase the temperature to $T=70$ K, which is lower than the critical temperature ($T_\mathrm{c}=80$ K), the magnetic Bragg peak moves to the lower $Q$-region ($q_\mathrm{m} \sim 0.27$ \r{A}$^{-1}$) and the smearing of the diffuse scatterings are apparent in the intensity map [Fig. \ref{fig:HRC_result_x0p6}(b)]. The fitting analysis in the constant-$E$ profile [Fig. \ref{fig:HRC_result_x0p6}(e)] show the coexistence of the Bragg and diffuse scatterings, in which the diffuse component is centered at $q_\mathrm{m} \sim 0.22$ \r{A}$^{-1}$. By integrating over the $Q$-range within those two components, the obtained constant-$Q$ profile [Fig. \ref{fig:HRC_result_x0p6}(h)] is well fitted by the summation of a resolution-limited Gaussian and a Lorentzian functions, implying that the static and fluctuating components coexist. When the temperature is above $T_\mathrm{c}$, the Bragg peak is disappeared in the intensity map [Fig. \ref{fig:HRC_result_x0p6}(c)]. The broad peak is observed in the constant-$E$ profile [Fig. \ref{fig:HRC_result_x0p6}(f)] and it has a Lorentzian shape as a function of $E$ [Fig. \ref{fig:HRC_result_x0p6}(i)], indicating that the diffusive spin fluctuations are realized. 

%\\[0.1in]

\section{MIEZE-NSE results at x = 0.6}
\label{AP:two}
\begin{figure*}[t]
\centering
\includegraphics[width=11 cm]{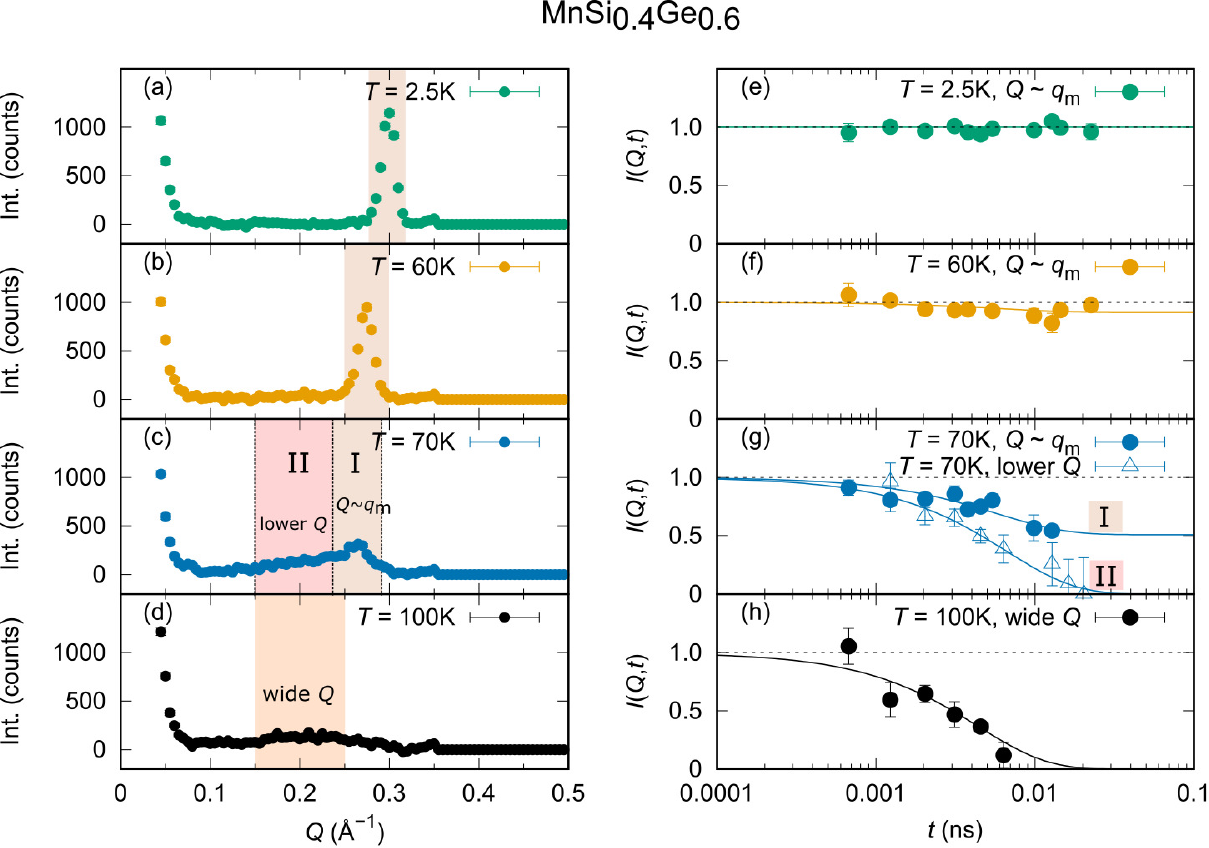}
\caption{MIEZE-NSE experimental results for $x = 0.6$. (a)-(d) $Q$-dependence of intensity at four representative temperatures, and (e)-(h) the intermediate scattering function of $I(Q,t)$ deduced from the selected $Q$-range given in (a)-(d).}
\label{fig:Iqt-comparison_x0p6}
\end{figure*}

We present the MIEZE-type neutron spin echo experimental results in the $x=0.6$ sample. Figures~\ref{fig:Iqt-comparison_x0p6}(a)-(d) are the intensity profiles as a function of $Q$ at four representative temperatures of $T=2.5,60,70,$ and $100$ K, the well-defined peaks attributed to the long-range orders are observed at low temperature and get broadened at high temperature. At $T=2.5$ K, the observed peak is clearly static since the $I(Q,t)$ profile is flat in time indicating that there is no fluctuating spins at all [Fig. \ref{fig:Iqt-comparison_x0p6}(e)]. In the middle range of the temperatures, we observed the coexistence of the static and fluctuating spin components. The weak fluctuating spin component is observed at $T=~60$~K [Fig. \ref{fig:Iqt-comparison_x0p6}(f)] where the static fraction slightly deviates from unity. At $T=70$ K, the fluctuating spin component gets stronger [Fig. \ref{fig:Iqt-comparison_x0p6}(g)]. We found two $I(Q,t)$ functions corresponding to the $Q$-ranges labeled I and II in Fig.~\ref{fig:Iqt-comparison_x0p6}(c). The $I(Q,t)$ profiles in the region I remains finite in the $t\rightarrow\infty$ limit, while in the region II decays to zero, suggesting the indication of the $Q$-dependence of the characteristic time $\tau$. The system becomes fully fluctuating at $T=100$~K [Fig. \ref{fig:Iqt-comparison_x0p6}(h)]. The critical temperature is obtained as $T_\mathrm{c} \sim 80$~K.

\bibliographystyle{apsrev4-2}
\bibliography{reference}

\end{document}